\documentclass[pre,twocolumn,floatfix]{revtex4}
\usepackage{amsmath}
\usepackage{graphicx}
\usepackage{color}
\usepackage{multirow}
\usepackage{array}

\newcommand{\moy}[1]{\langle #1 \rangle}

\def\NOTA#1{\vbox{\scriptsize\begin{quote}\textcolor{red}{\scshape #1}\end{quote}}}
\def\citeXXX{\marginpar{Ref needed}\textbf{[???]}}
\begin{document}
\title{Universal tree structures in directed polymers and models of
evolving populations}
\author{\'{E}ric Brunet}
\email{Eric.Brunet@lps.ens.fr}
\author{Bernard Derrida}
\email{Bernard.Derrida@lps.ens.fr}
\author{Damien Simon}
\email{Damien.Simon@lps.ens.fr}
\affiliation{Laboratoire de
Physique Statistique, \'Ecole Normale Sup\'erieure, 24, rue
Lhomond, 75231 Paris Cedex 05, France}

\begin{abstract}
By measuring or calculating coalescence times for several models of
coalescence or evolution, with and without selection, we show that the
ratios of these coalescence times become universal in the large size limit
and we identify a few universality classes.
\end{abstract}

\maketitle

Random trees appear in many contexts in biology, mathematics
and physics. In evolutionary biology, they represent the genealogies of
reproducing populations.
In physics, random trees appear in many systems 
such as DLA (diffusion limited aggregation)
\cite{wittensander}, coarsening, river networks
\cite{takayasu,cieplak.et.al}, diagrams in perturbation theory,
ultrametric structure of pure states in mean field spin glasses
\cite{MezardParisiVirasoro.87, BolthausenSznitman.98}, directed polymers
in a random medium \cite{KardarParisiZhang.86, Halpin-HealyZhang.95},
shocks in one-dimensional turbulence \cite{burgersgiraud, burgers1,
burgers2}, etc.

From a mathematical point of
view, one of the simplest examples of random trees is Kingman's coalescent \cite{kingman1,kingman2}: it
describes the coalescence tree of particles, where each pair of
particles has a probability $\delta t$ of coalescing into a single particle
during every infinitesimal time interval $\delta t$. The random tree
structures of Kingman's coalescent are identical to the genealogies
obtained in simple mean field models of neutral evolution such as the
Wright-Fisher model \cite{wrightWF,fisherWF}. In such models, each individual of
a population of fixed size $N$ at a given generation gives birth to
a random number of offspring and the population at the next generation is
obtained by choosing $N$ survivors at random among all these offspring. If
one follows the evolution over a large enough number of generations for
the initial condition to be forgotten, a steady state is reached where
the statistics of the genealogical tree of a large population are
identical to those of Kingman's coalescent.

Other random trees have been considered in the mathematical literature,
such as the $\Lambda$-coalescents \cite{Pitman.99,sagitov,Schweinsberg.00},
which generalize Kingman's coalescent and describe a wider class of
mean-field coalescence models \cite{varianceinfinie}. In the 
$\Lambda$-coalescent, each subset of $k$ particles among $ n $ particles
has a probability $\lambda_{n,k} \delta t$ of coalescing into a single
particle during an infinitesimal time $\delta t$. As a set of $n$ particles
can be considered as a subset of a larger set of $n+1$ particles, the
rates $\lambda_{n,k} $ have to satisfy some consistency relations:
the coalescence of $k$ particles in the subset of size $n$ happens in
two cases~: either these $k$ particles coalesce in the set of size
$n+1$ (rate $\lambda_{n+1,k}$) or they coalesce together with the
$(n+1)$-th particle (rate $\lambda_{n+1,k+1}$). Therefore
\begin{equation}
\lambda_{n,k}=\lambda_{n+1,k}+\lambda_{n+1,k+1}.
\end{equation}
This recursion leads to the following general expression for the
coalescence rates \cite{Pitman.99,sagitov}:
\begin{equation}
\label{eq:lambdarate:def}
\lambda_{n,k} = \int_0^1 x^{k-2} (1-x)^{n-k} \Lambda(x)\ dx,
\end{equation}
where $\Lambda$ is some positive measure on the interval $[0,1]$. With
these notations, Kingman's coalescent corresponds to
$\Lambda(x)=\delta(x)$. Another particular case, which has been studied, in
the context of spin glasses, is the Bolthausen-Sznitman coalescent
\cite{bolthausensznitman} for which $\Lambda(x)=1$. Trees in the
Kingman's coalescent and in the Bolthausen-Sznitman coalescent have
different statistical properties.

\begin{figure}[ht]
\includegraphics[width=4cm]{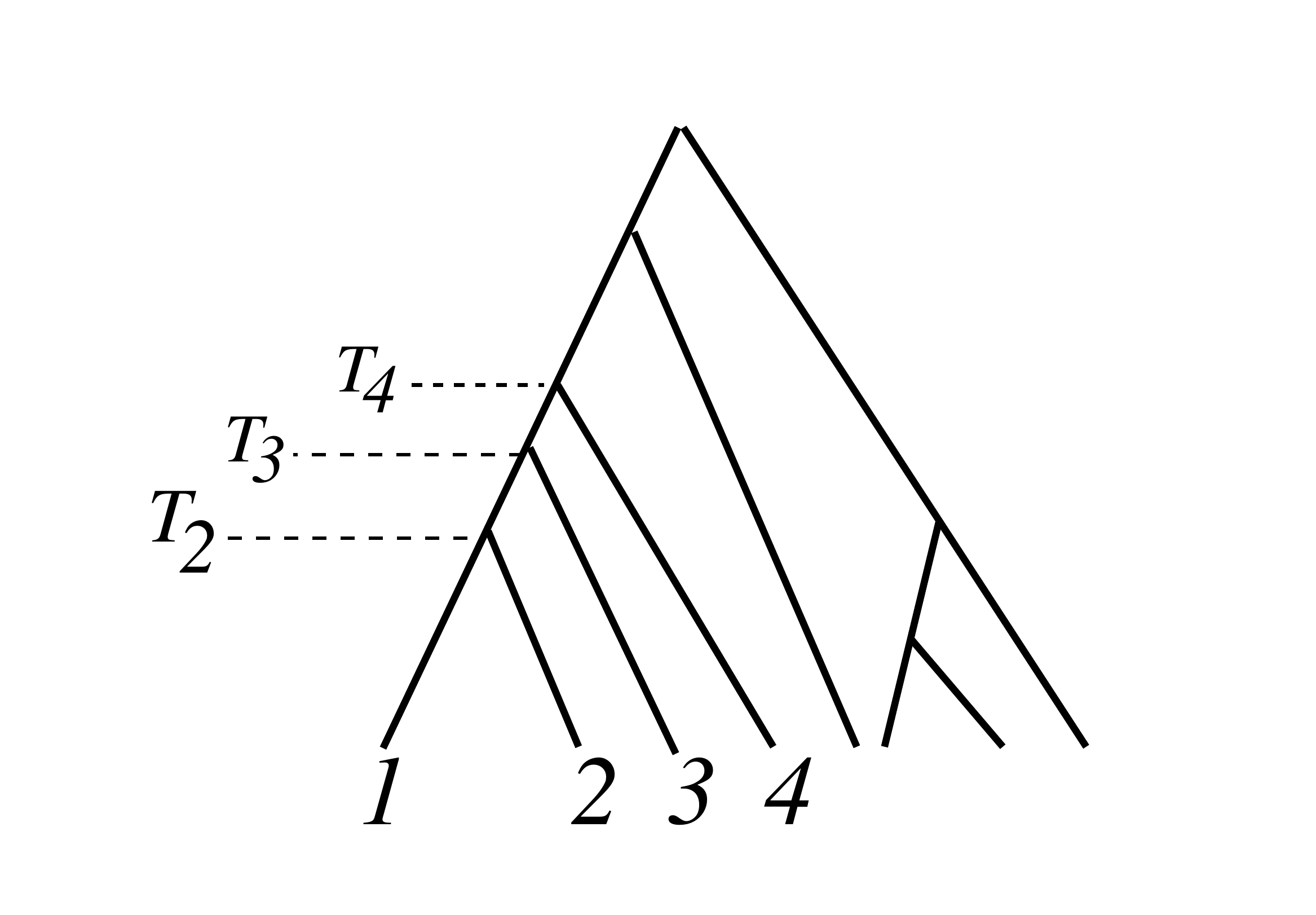}
\caption{The times $T_p$ are the ages of the most common ancestors of $p$
individuals chosen at random.}
\label{arbre}
\end{figure}

In order to compare different models of physical or biological systems which
generate random trees and to try to identify universality classes, we
consider here simple quantities characteristic of these random tree
structures. For a tree with a large number of end points, we define $T_p$
as the distance one has to go up into the tree to find the most recent
common ancestor of $p$ given points (see Fig.~\ref{arbre}). 

For models of evolving populations, the distance $T_p$ is the age of the
most recent common ancestor of $p$ individuals chosen at random in the
population. In general, it depends both on the generation at which these
$p$ individuals live, but also on the choice of the $p$ individuals, even
in the limit of very large trees. This double source of fluctuations for
the $T_p$ is reminiscent of what happens in mean field spin
glasses\cite{MezardParisiVirasoro.87}: as for the overlaps in Parisi's
theory, the distribution of the $T_p$ remains broad even when the size of
the population becomes very
large\cite{MezardParisiVirasoro.87,derridapeliti}.

For a given model, one can try to determine averages $\langle T_p \rangle$
or moments $\langle (T_p)^k \rangle$ of these times $T_p$ (the averages are
taken over all the branches of the tree, \textit{i.e.}\@ over all the population at
a given generation, and over all the random trees, \textit{i.e.}\@ over all the
generations in the language of models of evolution). In recent works
\cite{BDMM2.06,BDMM.07}, it was noticed that for a large class of mean field models of
evolution with selection, the ratios of these average times $\langle T_p
\rangle$ take, for a large population, simple universal values indicating that the genealogical
trees are distributed according to the statistics of the Bolthausen-Sznitman
coalescent. Theferefore, at the mean field level and for a large size of
the population, two universality classes seem to emerge for models of
evolution: Kingman's trees in the case of neutral evolution for which 
\begin{equation}
{\langle T_3 \rangle \over \langle T_2 \rangle} = {4 \over 3}
, \ \
{\langle T_4 \rangle \over \langle T_2 \rangle} = {3 \over 2}
, \ \
{\langle T_2 ^2\rangle \over \langle T_2 \rangle^2} =2 
, \ \
{\langle T_3 ^2\rangle \over \langle T_2 \rangle^2} = {26 \over 9},
\label{neutral}
\end{equation}
and Bolthausen-Sznitman's trees in the case of selection:
\begin{equation}
{\langle T_3 \rangle \over \langle T_2 \rangle} = {5 \over 4}
,\ \ 
{\langle T_4 \rangle \over \langle T_2 \rangle} = {25 \over 18}
, \ \
{\langle T_2^2\rangle \over \langle T_2 \rangle^2} = 2
, \ \
{\langle T_3^2\rangle \over \langle T_2 \rangle^2} = {11 \over 4}.
\label{selection}
\end{equation}

The goal of the present work is to try to measure these coalescence
ratios for other models of evolution, in particular to analyse the
effect of spatial fluctuations, and to argue that directed polymers in a
random medium are in the same universality classes as evolution
models in presence of selection.

The paper is organized as follows. In section~\ref{sec:neutral} we
consider, at the mean field level or in finite dimension,
coalescence models which are equivalent, as we will see, to neutral models
of evolution. Above two dimensions of space, the coalescence trees have the
same statistics \cite{limicsturm} as in mean field with coalescence times given by
Eq.~(\ref{neutral}), whereas in one dimension, they lead to a different
universality class for which we compute the ratios of coalescence times. In
section~\ref{sec:directed-polymers}, we consider the trees of optimal paths
in the problem of directed polymers in a random medium. Our numerical
results will show that at the mean field level, the trees satisfy the
Bolthausen-Sznitman statistics Eq.~(\ref{selection}), whereas the ratios of
coalescence times vary with dimension as expected by the known universality
classes of the problem.

\section{Coalescence and models of neutral evolution }
\label{sec:neutral}

\subsection{Kingman's coalescent}

Kingman's coalescent \cite{kingman1,kingman2} is a mean field model of coalescing
particles: during each infinitesimal time interval $\delta t$ every pair of
particles has a probability of coalescing into a single particle. Therefore
if one starts with $p$ particles, there is a random waiting time $\tau_p$
until a coalescence event occurs when these $p$ particles become $p-1$
particles. Then there is another random time $\tau_{p-1}$ until a pair
among these $p-1$ particles coalesce (and one is left with $p-2$
particles),
and so on. The times $\tau_k$ are independent and distributed according to
exponential distributions
\begin{equation}
 \rho_k(\tau_k) = \frac{k(k-1)}{2} \exp\left( - \frac{k(k-1)}{2}
\tau_k\right),
\label{rhok}
\end{equation}
and the time $T_p$ for $p$ particles chosen at random to coalesce is given by:
\begin{equation}
 T_p = \tau_p+\tau_{p-1}+\cdots + \tau_3+\tau_2.
\end{equation}
This allows one to recover easily the values of Eq.~(\ref{neutral}).
In fact the whole generating functions of the times $T_p$ can be calculated:
\begin{equation}
\left\langle e^{\lambda T_p} \right\rangle = 
\prod_{k=2}^p { k(k-1) \over k(k-1) - 2 \lambda} \label{elambdaT}.
\end{equation}
In particular one can notice that the time $T_2$ has an exponential
distribution.

\subsection{Wright-Fisher model}

The Wright-Fisher model \cite{wrightWF,fisherWF} is one of the simplest neutral models of an
evolving population. It describes a population of constant size $N$ with
non-overlapping generations and asexual reproduction. At each generation,
all the population is replaced by $N$ new individuals with the following
rule: each individual at a given generation has its parent randomly chosen
among the $N$ individuals at the previous generation. If one goes backward
in times, the lineage of an individual performs a random walk on a fully
connected graph of $N$ sites. Following the lineages of $p$ individuals is
the same as following $p$ coalescing random walks on this fully connected
graph. Since the random walks are independent, the statistics of
coalescence times can be easily calculated \cite{kingman1,kingman2}:
for $p$ individuals chosen at random at generation $g$, the time $T_p$ is
the age of their most recent common ancestor, \emph{i.e.}\@ $T_p$ is the number of time
steps for the $p$ random walkers on the fully connected graph to coalesce.
At each generation in the past, two distinct lineages have a probability
$1/N$ of merging, thus $T_2$ scales as the size $N$ of the population. For
fixed $p >2$, the probability that a pair of lineages coalesce is
$1/N$ whereas multiple coalescences occur with higher powers
of $1/N$ for large $N$. One can then neglect these multiple
coalescences and
\begin{equation}
T_p \simeq N (\tau_p + \tau_{p-1} + \cdots + \tau_3+ \tau_2),
\end{equation}
where the times $\tau_k$ are distributed according to Eq.~(\ref{rhok}),
implying that the statistics of the times $T_p$ are exactly the same as
in Kingman's coalescent Eq.~(\ref{neutral}).

\subsection{Coalescing random walks in finite dimension}
\label{neutralmodel}

We are now going to look at coalescing random walks on an hypercube of
$N=L^d$ sites in dimension $d$ with periodic boundary conditions. We
consider the continuous time case, where during infinitesimal time interval
$\delta t$, each walker on the hypercube has a probability $\delta t$ of
hopping to each of its neighboring sites, and whenever two walkers occupy
the same site, they instantaneously coalesce into a single walker. If
$T_2(\vec{r})$ is the coalescence time between two walkers at a distance
$\vec{r}$ apart, its evolution is
\begin{equation}
T_2(\vec{r}) =\begin{cases} \delta t + T_2(\vec{r}) & \text{with probability $1 -
4 d\, \delta t$},
\\ \delta t + T_2(\vec{r} + \vec{e_i}) &\text{with probability
$2\delta t$},
\end{cases}
\label{first-passage}
\end{equation}
where $\vec{e_i}$ is one of the $2d$ unit vectors on the hypercubic
lattice.

It is clear that the distance between the two walkers performs a random
walk and that $T_2$ is simply the first time that this distance vanishes.
This is of course a very well known first passage problem
\cite{Montroll.64,Montroll.65} which
can be solved easily (it reduces to the inversion of a Laplacian): the
generating function of $T_2(\vec{r})$ satisfies for $\delta t \ll 1$ and for
$\vec{r} \neq 0$
\begin{multline}
\left\langle e^{\lambda T_2(\vec{r})} \right\rangle 
=e^{\lambda \delta t}\Bigg[(1 - 4 d\, \delta t) \left\langle e^{\lambda T_2(\vec{r})} \right\rangle 
\\+ 2 \delta t \sum_{i=1}^{2 d} \left\langle e^{\lambda T_2(\vec{r} + \vec{e_i})}
\right\rangle \Bigg], 
\label{eT2}
\end{multline}
where $\langle \cdot \rangle$ denotes an average over all the random walks. At $\vec{r}=0$, it satisfies the boundary condition 
\begin{equation}\left\langle e^{\lambda
T_2(\vec{0})} \right\rangle=1.
\label{T2(0)}
 \end{equation}
For $\vec r\ne\vec0$, one can rewrite Eq.~(\ref{eT2}) as
\begin{equation}
 \lambda \left\langle e^{\lambda T_2(\vec{r})} \right\rangle 
+ 2 \sum_{i=1}^{2 d} \left[\left\langle e^{\lambda T_2(\vec{r} + \vec{e_i})}
\right\rangle
-\left\langle e^{\lambda T_2(\vec{r})} \right\rangle \right] = 0,
\label{eT2bis}
\end{equation}
and this can be easily solved in Fourier space to give
\begin{equation}
\left\langle e^{\lambda T_2(\vec{r})} \right\rangle = 
A(\lambda) \sum_{n_1=0}^{L-1}
\cdots
\sum_{n_d=0}^{L-1}
 {\exp{ {2 i \pi \vec{n}. \vec{r} \over L}} \over \lambda + 4 \sum_{i=1}^d (
\cos {2 \pi n_i \over L} - 1)}
,
\end{equation}
where the constant $A(\lambda)$ is fixed by the condition of Eq.~(\ref{T2(0)}).

Starting with two particles at random positions on the lattice and averaging over these two positions leads to 
\begin{equation}
\left\langle e^{\lambda T_2} \right\rangle = 
\left[ 1 + \sum_{\vec{n} \neq 0} { \lambda \over \lambda + 4 \sum_{i=1}^d (
\cos {2 \pi n_i \over L} - 1)}
 \right]^{-1}. 
\label{T2gen}
\end{equation}
This implies that
\begin{gather}
\label{T2a}
\langle T_2\rangle = \sum_{\vec{n} \neq 0} { 1 \over 4 \sum_{i=1}^d (1- 
\cos {2 \pi n_i \over L} )},
\\
\begin{split}
\langle T_2^2\rangle &= 2 \left[\sum_{\vec{n} \neq 0} { 1 \over
4 \sum_{i=1}^d (1- 
\cos {2 \pi n_i \over L} )} \right]^2 \\
&\quad+2 \sum_{\vec{n} \neq 0} \left[ { 1 \over 4 \sum_{i=1}^d (1-
\cos {2 \pi n_i \over L} ) } \right]^2 .
\end{split}
\label{T2b}
\end{gather}

For large $L$, it is well known \cite{Montroll.64,Montroll.65} that Eq.~(\ref{T2a}) gives
\begin{equation}
\label{eq:timescales}
\langle T_2 \rangle \propto
\begin{cases}
 L^2 & \text{for $d=1$}, \\
 L^2 \ln L & \text{for $d=2$}, \\
 L^d & \text{for $d > 2$}.
\end{cases}
\end{equation}
On the other hand, one can show that the second term in the
right-hand-side of Eq.~(\ref{T2b}) grows
as $L^d$ in dimension $d> 4$ and as $L^4$ in dimension $d< 4$. Therefore for
$d \geq 2$, the ratio $ {\langle T_2^2 \rangle / \langle T_2 \rangle^2}$
goes to 2
when $L \to \infty$, as in the mean field case Eq.~(\ref{neutral}).

In fact it has been proved \cite{cox,limicsturm} that in $d \geq 2$ (and for large $L$)
 the whole genealogies of $p$ individuals (averaged
over all their positions) are given by the Kingman coalescent, up to the
rescaling (\ref{eq:timescales}). In particular the distribution of the time
$T_2$ is exponential.

\subsection{Coalescing random walks in one dimension}

In dimension $d< 2$, the two terms in the right-hand-side of
Eq.~(\ref{T2b}) are comparable, and the ratio $ {\langle T_2^2 \rangle /
\langle T_2 \rangle^2} $  no longer converges to $2$.

In dimension $d=1$ the calculation of all the moments of the times $T_p$ is rather straightforward.
First one can easily solve Eq.~(\ref{eT2bis}) for periodic boundary
conditions with condition Eq.~(\ref{T2(0)}) and one gets
\begin{equation}
\left\langle e^{\lambda T_2(r) } \right\rangle 
=
{\left({4 - \lambda + \sqrt{\lambda^2 - 8 \lambda} \over 4} \right)^{{L \over 2}-r}
+\left({4 - \lambda + \sqrt{\lambda^2 - 8 \lambda} \over 4} \right)^{r-{L \over 2} } \over 
 \left({4 - \lambda + \sqrt{\lambda^2 - 8 \lambda} \over 4} \right)^{{L \over 2}}
+\left({4 - \lambda + \sqrt{\lambda^2 - 8 \lambda} \over 4} \right)^{-{L \over 2} } }.
\end{equation}
For large $L$, this becomes a scaling function of $\lambda L^2$ and of $r/L$
\begin{equation}
\left\langle e^{\lambda T_2(r) } \right\rangle 
\simeq{ \cos {(L-2 r) \sqrt{\lambda} \over 2\sqrt2}
 \over \cos  {L \sqrt{\lambda}\over2\sqrt2 }}
= {\sin {r\sqrt{\lambda}\over{\sqrt2}} 
+ \sin {(L-r)\sqrt{\lambda}\over\sqrt{2}} \over
\sin {L\sqrt{\lambda}\over\sqrt{2}} }
\label{T2r}
\end{equation}
and, averaging over $r$, one gets
\begin{equation}
\left\langle e^{\lambda T_2 } \right\rangle 
\simeq {2\sqrt2 \over L \sqrt{\lambda} }\tan { L \sqrt{\lambda} \over
2\sqrt2},
\label{T21d}
\end{equation}
which shows that the distribution of $T_2$ is no longer exponential.

One can write down the equations satisfied by the generating functions of
the times $T_p$. For large $L$ and $\lambda= {\cal O}(L^{-2}) $ the solution is
\begin{equation}
\left\langle e^{\lambda T_p(r_1,\ldots,r_p) } \right\rangle \simeq \sum_{k=1}^p
 {\sin ( r_k \sqrt{\lambda/2} )
 \over \sin (L \sqrt{\lambda/2} ) },
\label{eq21}
\end{equation}
where the $r_k$ are the distances between consecutive particles along the
ring (one has of course $r_1+\cdots+ r_p=L$). In particular, for $p=2$,
$r_1=r$ and $r_2=L-r$, one recovers Eq.~(\ref{T2r}).
Averaging Eq.~\eqref{eq21} over all the positions
of the $p$ particles on the ring leads to
\begin{equation}
\left\langle e^{\lambda T_p } \right\rangle \simeq p ( p-1) 
\int_0^1 dx\  
 {\sin ( L x \sqrt{\lambda/2} )
 \over \sin (L \sqrt{\lambda/2} ) }
(1-x)^{p-2}.
\label{Tp1d}
\end{equation}

From Eq.~(\ref{Tp1d}) one can then obtain all the moments of $\langle T_p \rangle$. For example, one has
\begin{equation}
\langle T_p \rangle \simeq {(p-1)(p+4) \over 12 (p+1)(p+2)} L^2
\end{equation}
and one can show
\begin{equation}
{\langle T_3 \rangle \over \langle T_2 \rangle} = {7 \over 5}
, \ 
{\langle T_4 \rangle \over \langle T_2 \rangle} = {8 \over 5}
, \ 
{\langle T_2 ^2\rangle \over \langle T_2 \rangle^2} ={12 \over 5}
, \ 
{\langle T_3 ^2\rangle \over \langle T_2 \rangle^2} = {124 \over 35},
\label{1d}
\end{equation}
in contrast with Eq.~(\ref{neutral}) and Eq.~(\ref{selection}).

One could repeat the calculations which lead to
Eqs.~(\ref{T2a},~\ref{T2b},~\ref{eq:timescales}) and Eq.~(\ref{1d}) for models
of coalescence on other lattices or with more general jumping rates. As
long as the motion of the coalescing particles remains diffusive, one
would recover the same values Eq.~(\ref{neutral}) or Eq.~(\ref{1d}) for
the statistics of the trees.

\subsection{Neutral evolution in finite dimension}

One can try to generalize the Wright-Fisher model to the finite
dimensional case, for example by considering an hypercube with a finite
population of fixed size $m$ on each lattice site, and the case where each
individual chooses its parent in the previous generation with a
probability $p$ on the same lattice site and with probability $1-p$ on
one of the neighboring sites. The study of the genealogies in this case
is obviously the same problem as following the coalescences of the
lineages which perform random walks on this lattice. Therefore in
dimension $d=2$ and above, the trees are given by the statistics
Eq.~(\ref{neutral}) of Kingman's coalescent whereas in dimension
$d=1$ they will be in the universality class Eq.~(\ref{1d}) of coalescing
random walks in one dimension.

\section{Directed polymers in a random medium}
\label{sec:directed-polymers}

Directed polymers in a random medium is one of the simplest examples of
a strongly disordered system \cite{Kardar.87,KardarZhang.87, Halpin-HealyZhang.95,
EmigKardar.01}. It
describes directed paths in a random energy landscape. In its zero
temperature version, the problem reduces to finding the optimal path,
\textit{i.e.}\@ the path of minimal energy in this random energy
landscape. The optimal paths starting at the same point but arriving at different points give rise to a
tree structure, that we try to characterize in this section by measuring
the coalescence times $T_p$.

\begin{figure}[!ht]
\centering
\raisebox{-\height}{\includegraphics[width=.44\columnwidth]{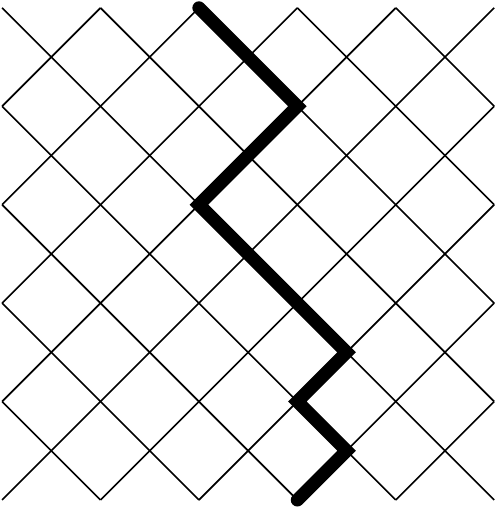}}
\ 
\raisebox{-\height}{\includegraphics[width=.44\columnwidth]{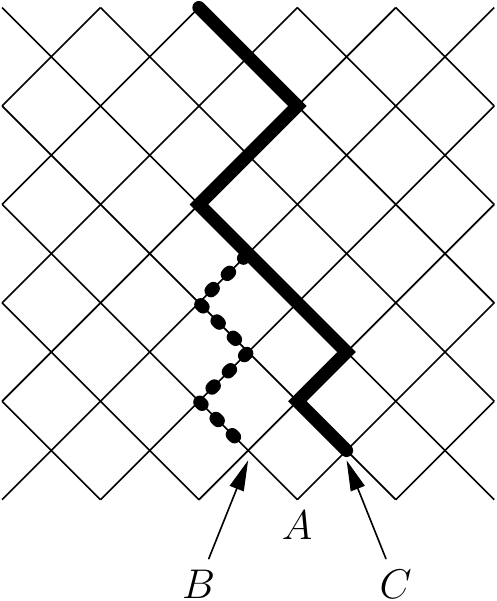}}
\caption{\textit{(Left)} a directed polymer in dimension $1+1$. The
``time'' direction is vertical. \textit{(Right)} 
A directed polymer arriving at $A$ comes either
from $B$ or from $C$, whichever is more energetically favorable:
In the example shown, the coalescence time of the directed polymers
arriving at $B$ and $C$ is four.}
\label{dpfig}
\end{figure}

\begin{figure}[!ht]
\centering
\includegraphics[width=.8\columnwidth]{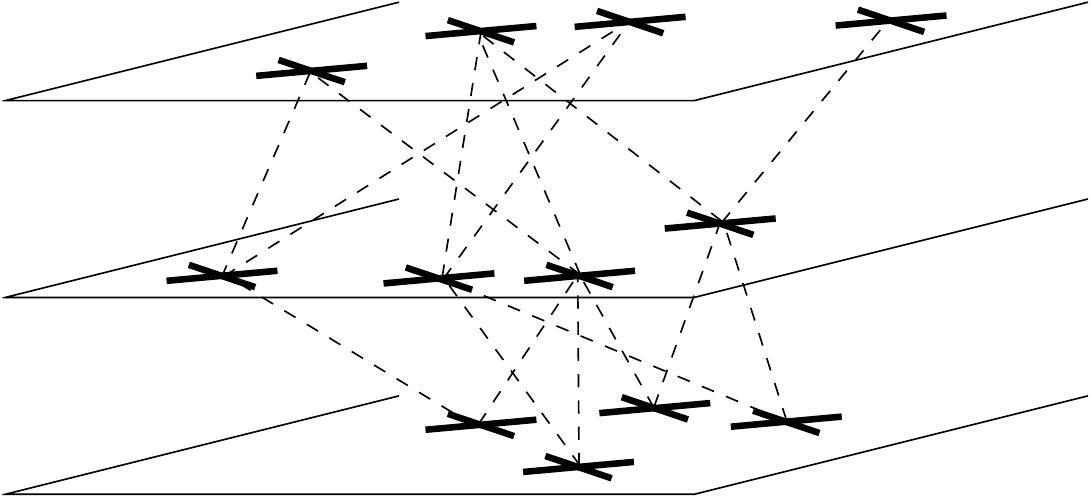}
\caption{Directed polymer in mean-field. At a given time, each of the
$N=4$ sites is connected to $M=2$ random sites at the previous time.}
\label{dpMF}
\end{figure}

A directed polymer in dimension $d+1$ is a line extending in one of the 
directions (traditionnaly called ``time'', and which we represent as
the vertical direction in Fig.~\ref{dpfig} and Fig.~\ref{dpMF})
with some random excursions in the $d$ other transverse directions
(see Fig.~\ref{dpfig}).

We consider here directed polymers on a lattice which is infinite in the
``time'' direction but finite and periodic in the $d$ transverse
directions. In
each time section, there are $N=L^d$ sites located on a $d$-dimensional
hypercube of linear size $L$ with periodic boundary conditions.
Each site in a given time section is connected to $M=2^d$ sites in the
previous time section (and it is also connected to $M$ other sites in the
next time section).
The way each site is connected is shown for dimension $1+1$ in
Fig.~\ref{dpfig}. In higher dimension, we generalized the lattice of
Fig.~\ref{dpfig} in the following way: let $\vec x=(x_1,x_2,\ldots,x_d)$ be the
transverse coordinates of a given site; the $x_i$ are
integers at even times and half-integers at odd times, and the $M=2^d$
potential parent sites² of $\vec x$ have coordinates
$(x_1\pm1/2,x_2\pm1/2,\ldots, x_d\pm1/2)$ in the previous time
section.

We consider also a mean-field version (Fig.~\ref{dpMF}), where there is no spatial structure in the transverse directions: A time section consists
of a set of $N$ sites, and each of them is connected to $M$
sites chosen at random among the $N$ sites of the previous time section, where $M$ might
be any number between 2 and $N$. 

We assume that each link $(AB)$ between two connected sites $A$ and $B$
carries a random energy $\epsilon_{(AB)}$. The energy $E$ of the polymer
is then the sum of all the energies $\epsilon_{(AB)}$ of the visited links.

We choose an origin where the polymer starts, and for any given site $A$
on the lattice, we call $E_A$ the minimal energy of the polymer over
all the possible directed paths connecting this origin to $A$. At zero temperature, the directed polymer chooses
the path which minimizes its energy and one has the simple recursion
relation
\begin{equation}
E_A=\min\Big(E_B+\epsilon_{(AB)}, E_C+\epsilon_{(AC)}, \cdots\Big),
\label{DPrec}
\end{equation}
where $B$, $C$, \ldots\ are the $M$ potential parent sites of site $A$.

For any pair of sites $A$ and $A'$ in the same time section, we define their
coalescence time (see Fig.~\ref{dpfig}) as the number of up steps during which the two optimal paths
arriving at $A$ and $A'$ differ (we suppose that the origin of the
directed polymers is at a remote enough time in the past for the paths to
coalesce).
In a similar way, we define the coalescence times of any group of $p$
different sites as the maximal coalescence time of any pair within the
$p$ sites. All these quantities depend on the chosen sites and on the
realization of the disorder, and, as in the previous section, we note by $\langle\cdot\rangle$ the
average over the choice of sites and the disorder.
In this section, we consider the averaged coalesence time  $\langle T_p
\rangle$ and the averaged square of the coalesence
time $\langle T_p^2 \rangle$  of $p$ sites. 

\begin{figure}[!ht]
\centering
\includegraphics[width=.8\columnwidth]{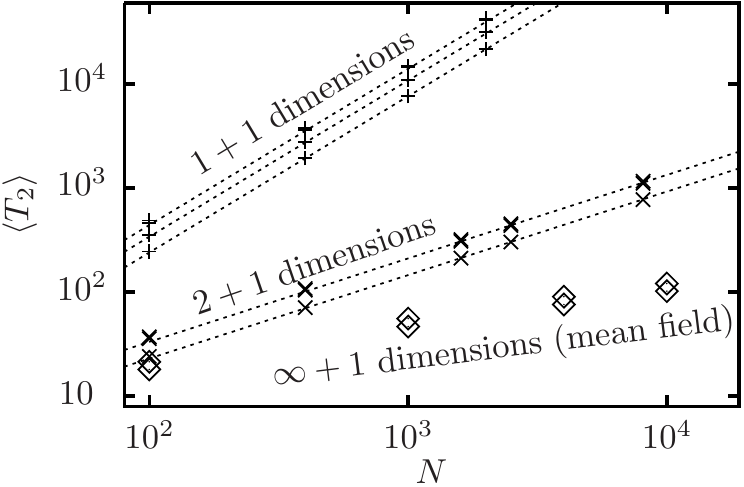}
\caption{Averaged coalescence time $\langle T_2\rangle $ of two
individuals for several models of directed polymers in dimensions $1+1$ and
$2+1$, and in mean-field, as a function of the number $N$ of sites in
each time section.
The data are compared  to the prediction $\langle
T_2 \rangle \propto N^{1/(d\nu)}$ in dotted lines for dimensions $1+1$ and $2+1$. Note
that, by chance, two out of the four models in dimension $1+1$ and two out
of the three models in dimension $2+1$ have nearly the same prefactor and
their data are undistinguishable.}
\label{T2_zerotemp}
\end{figure}

We have simulated four models in dimension $1+1$; from top to bottom on
Fig.~\ref{T2_zerotemp}:
\begin{itemize}
\item on the lattice of Fig.~\ref{dpfig} with a discrete distribution of $\epsilon$ with
values $\epsilon=0$ or $\epsilon=1$ with probabilities $1/2$,
\item on the lattice of Fig.~\ref{dpfig} with a uniform distribution of $\epsilon$ in
$[0,1]$,
\item on the lattice of Fig.~\ref{dpfig}, with negative values of $\epsilon$, distributed according to $\rho(\epsilon)=e^{\epsilon}\,\theta(-\epsilon)$ 
\item on a square lattice where each site is connected to $M=3$ parents
(just above itself, on its right and on its left)
where $\epsilon$ takes \emph{positive} values, with an exponentially
\emph{decreasing} distribution:
$\rho(\epsilon)=e^{-\epsilon}\,\theta(+\epsilon)$.
\end{itemize}
In dimension $2+1$, we have simulated three models
all on the lattice with $M=4$ ancestors described above; from top to bottom on
Fig.~\ref{T2_zerotemp}:
\begin{itemize}
\item with an exponentially \emph{increasing} distribution:
$\rho(\epsilon)=e^{+\epsilon}\,\theta(-\epsilon)$
\item with a uniform distribution of $\epsilon$ in $[0,1]$,
\item with an exponentially \emph{decreasing} distribution:
$\rho(\epsilon)=e^{-\epsilon}\,\theta(+\epsilon)$.
\end{itemize}
Finally, we have simulated two models in mean-field with a uniform distribution of $\epsilon$ in
$[0,1]$ and either $M=2$ or $M=4$ random ancestors for each site ($M=2$
is above $M=4$ in Fig.~\ref{T2_zerotemp}.).
Our data for all these models are plotted together with the same symbol for
each dimension to emphasize the universality of our results.

To measure the $T_p$'s, the conceptualy simplest way is to update a
$N\times N$ matrix
containing for all pairs $(i,j)$ of individuals the time $T_2(i,j)$ of
their most commun ancestor. Indeed, for an
arbitrary number $p$ of individuals, one has
$T_p(i_1,\ldots,i_p)=\max[T_2(i_1,i_2), T_2(i_1,i_3),\ldots,
T_2(i_1,i_p)]$, so that the matrix of the $T_2$'s contains all the relevant
information. Updating this matrix at each time step is easy: The $T_2$ of
two different sites is one plus the $T_2$ of their parents, and the $T_2$
of a site with itself is zero.
Because updating at each time step a $N\times N$ matrix is time consuming, we
used a more sophisticated method\cite{BDMM.07} where we keep track of the genealogical
tree of all the sites at a current time: there are of course $N$ sites at
the current time, and at most $N-1$ nodes, where a node is a site from
previous times which is the most recent common ancestor of two sites at the
current time. At each time step, updating the whole tree takes a time
linear in $N$, and averaging the $T_p$ over all the choices of $p$
individuals takes also a time linear in $N$, 
as one simply has to recursively walk down the tree from its root and
count for each node the number of times it is the most recent commun ancestor
of $p$ sites in the current time. This algorithm is described in more
details in \cite{BDMM.07}.

For each data point, we have run one long simulation and 
averaged our results over all the time steps once the steady state was
reached. This is equivalent to averaging over many independent
realizations if we run a simulation for a time much longer than
the correlation time, which we estimated to be of the order of magnitude of $\langle T_2
\rangle$. All of our simulations were at least 20\,000
times longer than $\langle T_2 \rangle$.

In Fig.~\ref{T2_zerotemp}, we plot the coalescence time  $\langle
T_2\rangle$ as a function of the system size. For directed polymers on a
lattice which is infinite both in the time direction and in the $d$
transverse directions, the transverse displacement of the
optimal path scales like $t^\nu$, where $t$ is the length of the directed
polymer and $\nu$ is a universal exponent\cite{Halpin-HealyZhang.95}
equal to $\nu_{1+1}=2/3$ in dimension $1+1$ and $\nu_{2+1}\approx0.624$
in dimension $2+1$.
In our setup, with a lattice finite of linear
size $L$ in the transverse directions, this
scaling can only hold as long as $t<T_\text{corr}$ with
$T_\text{corr}^\nu=L=N^{1/d}$. This time $T_\text{corr}$ 
is the
correlation time on the scale of which the system
forgets its initial condition. Moreover, if we consider several sites
and the optimal paths arriving at these sites, these paths coalesce on a
time scale of the order of $T_\text{corr}$, as can be seen on
Fig.~\ref{T2_zerotemp}.

In mean-field with a finite number $M$ of potential ancestors per site, there is no
notion of distance in the transverse directions, and the exponent $\nu$ is
meaningless. We therefore expect a different scaling.
The problem of zero-temperature mean-field directed polymers can be
formulated\cite{BrunetDerrida.04} as a noisy
Fisher-KPP like 
equation \cite{Fisher.37, KPP.37}. Recently, a phenomenological theory of
coalescence trees in models of Fisher-KPP fronts
suggested\cite{BDMM.07,BDMM.06,BrunetDerrida.97} that the coalescence
time in such models should be of order $T_\text{corr}\propto(\ln N )^3$.
On Fig.~\ref{T2_zerotemp}, one can see that the data seem
to have a slower growth than a power law, but the values of $N$ we
simulated here are too small to check the $(\ln N )^3$ prediction.
Better simulations on a closely related model are presented
in~\cite{BDMM.07} where the $(\ln N )^3$ scaling appears clearly.

\begin{figure}[!ht]
\centering
\includegraphics[height=.44\columnwidth]{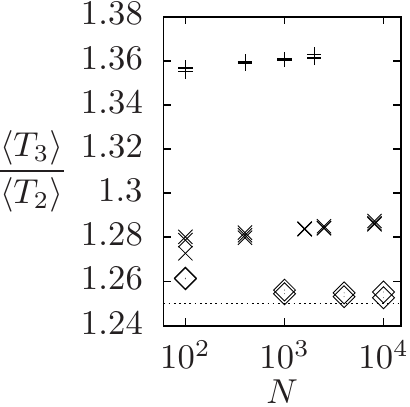}
\ 
\includegraphics[height=.44\columnwidth]{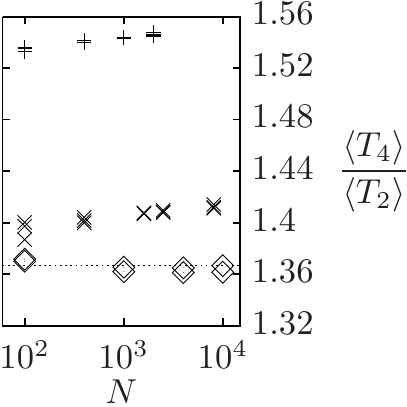}
\caption{Ratios of coalescence times for directed polymers at zero
temperature as a function of the size $N$ of the system in, from top to
bottom, dimension $1+1$, dimension $2+1$ and mean-field. The dotted line
represents the prediction Eq.~\eqref{selection} for mean-field in the limit
of infinite size. \textit{(Left)} ratios $\langle T_3\rangle /\langle
T_2\rangle $. \textit{(Right)} ratios $\langle T_4\rangle /\langle
T_2\rangle $.}
\label{ratios_DPT0}
\end{figure}

We now turn to the ratios of coalescence times. Fig.~\ref{ratios_DPT0}
shows the ratios $\langle T_3\rangle /\langle T_2\rangle $ and $\langle
T_4\rangle /\langle T_2\rangle $ as a function of the system size for all
the models we study (four models in dimension $1+1$, three in dimension
$2+1$ and two in mean-field). Numerically, for large~$N$, these ratios
seem to depend only on the dimension, and not on the
distribution $\rho(\epsilon)$ of the bond energies, nor on the shape of the lattice. The results in mean-field are
compatible with the prediction that for an infinitely large system in the
Fisher-KPP front equation class\cite{BDMM.07}, the genealogical tree
converges to a Bolthausen-Sznitman coalescent, with ratios given by
Eq.~\eqref{selection}. In dimensions $1+1$ and $2+1$,
our numerical results indicate clearly that we have tree statistics
different from the Bolthausen-Sznitman coalescent, and also different
from the Kingman coalescent for which $\langle
T_3\rangle /\langle T_2\rangle$ would be $3/2$ as in Eq.~\eqref{neutral}.
\begin{figure}[!ht]
\includegraphics[height=.44\columnwidth]{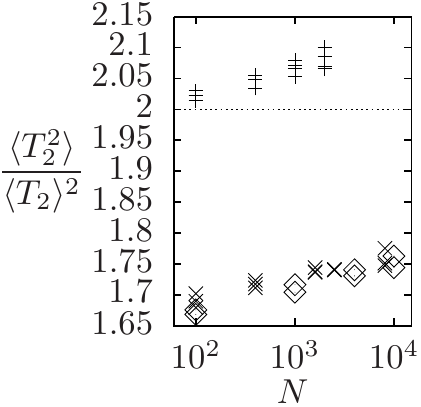}
\ 
\includegraphics[height=.44\columnwidth]{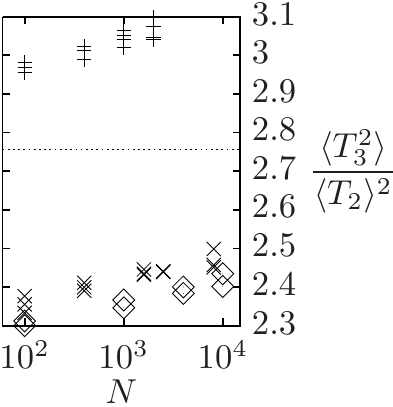}
\caption{
Ratios of moments of the coalescence times for directed polymers at zero
temperature as a function of the size $N$ of the system in, from top to
bottom, dimension $1+1$, dimension $2+1$ and mean-field. The dotted line
represents the prediction Eq.~\eqref{selection} for mean-field in the
limit
of infinite size. \textit{(Left)} ratios $\langle T_2^2\rangle /\langle
T_2\rangle^2 $. \textit{(Right)} ratios $\langle T_3^2\rangle /\langle
T_2\rangle^2 $.}
\label{ratios2_DPT0}
\end{figure}

On Fig.~\ref{ratios2_DPT0}, we show the ratios $\langle T_2^2\rangle
/\langle T_2\rangle^2 $ and $\langle
T_3^2\rangle /\langle T_2\rangle^2 $. Here, the situation is less
clear: the symbols for the different
models do not superpose and the
ratios do not seem to have converged (in particular, the mean-field
ratios are rather far from the prediction Eq.~\eqref{selection}). For some
reason we do not understand, it seems that the $\langle
T_p^2\rangle/\langle T_2\rangle^2 $ need much larger values of $N$ to
converge to their final values than the $\langle T_p\rangle/\langle
T_2\rangle$. We already observed a similar phenomenon on an exactly solvable
related model\cite{BDMM.07}. 

We also measured the ratios $\langle T_N\rangle/\langle T_2\rangle$,
where $T_N$ is the age of the most recent common ancestor of the whole
population and found these ratios to be close to 1.93 in dimensions $1+1$
and $2+1$, while it diverges in mean-field.

\subsection{Long tail distributions}

\begin{figure}[!ht]
\centering
\includegraphics[width=.8\columnwidth]{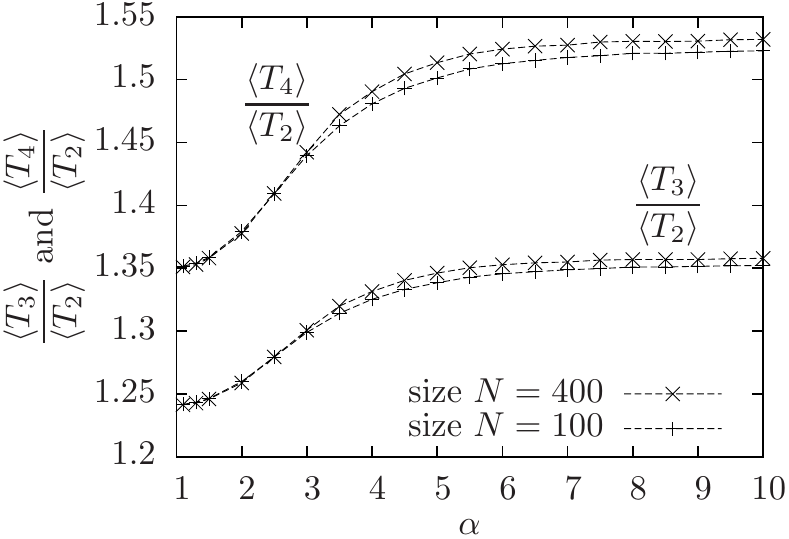}
\caption{Ratios $\langle T_3\rangle/\langle T_2\rangle$
and $\langle T_4\rangle/\langle T_2\rangle$ as a function of the
exponent $\alpha$ appearing in the noise Eq.~\eqref{longtail}, for two
different system sizes in dimension $1+1$.}
\label{fig1tail}
\end{figure}

In the directed polymer problem, it is known that the scaling regime
is modified when the distribution $\rho(\epsilon)$ of the energies of the
bonds decays as a power law $\rho(\epsilon)\propto |\epsilon|^{-\alpha}$
for large negative $\epsilon$: when $\alpha<\alpha_c$ with $\alpha_c\simeq 7$,
the directed polymer in dimension $1+1$ has an anomalous scaling \cite{Halpin-HealyZhang.95,Zhang.90} and the
exponent $\nu$ depends on $\alpha$.
We have measured the coalescence times
in dimension $1+1$ for a distribution of energies given by
\begin{equation}
\rho(\epsilon)= {A(\alpha)\over(1+|\epsilon|)^\alpha},
\label{longtail}
\end{equation}
with $\alpha>1$ and $A(\alpha)$ such that $\rho(\epsilon)$ is normalized,
for sizes $N=100$ and $N=400$. The ratios $\langle T_3\rangle/\langle
T_2\rangle$ and $\langle T_4\rangle/\langle T_2\rangle$ are presented in
Fig.~\ref{fig1tail}.
We observe that, for large $\alpha$, these ratios
converge towards the universal values shown on
Fig.~\ref{ratios_DPT0}, while for $\alpha\to1^+$, they seem to 
converge close to, respectively, $1.24$ and $1.35$.

\begin{figure}[!ht]
\centering
\includegraphics[width=.8\columnwidth]{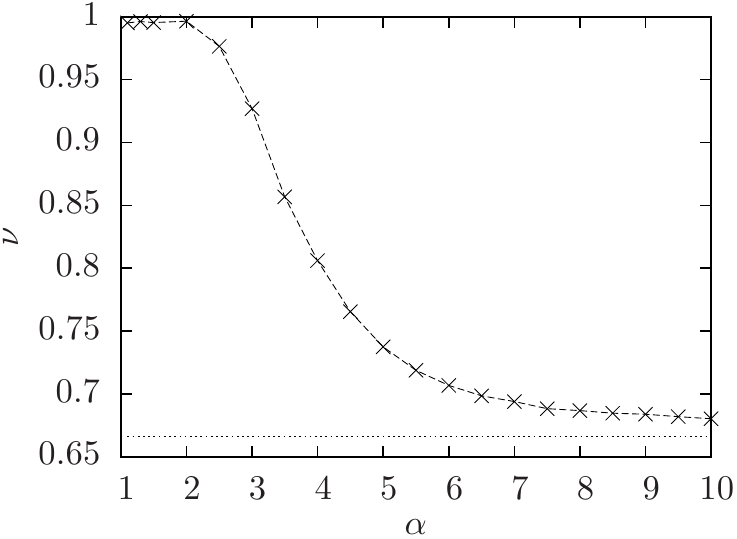}
\caption{Estimate of the exponent $\nu$ of the directed polymer as a function of
the exponent $\alpha$ appearing in the distribution of $\epsilon$ Eq.~\eqref{longtail}. This
exponent has been evaluated from the formula $\ln(4)/\ln\big[{\langle
T_2{\scriptstyle (N=400)}\rangle}/{\langle
T_2{\scriptstyle(N=100)}\rangle}\big]$. The universal value $\nu_{1+1}=2/3$
for distributions decaying fast enough is also shown.}
\label{fig2tail}
\end{figure}

As we expect $\langle T_2\rangle$ to scale like $N^{1/\nu(\alpha)}$, it is
possible to obtain a rough estimate of the exponent $\nu(\alpha)$  from
the only two datapoints at sizes $N=100$ and $N=400$. This estimate
is shown on Fig.~\ref{fig2tail}.
The exponent $\nu(\alpha)$ seems to converge toward the universal value
$\nu_{1+1}=2/3$ for large $\alpha$, while it seems to be $1$ for
$1<\alpha\lesssim 2$.

As with previous numerical studies\cite{Halpin-HealyZhang.95},
our results are not precise enough to determine precisely the
critical $\alpha_c$ above which $\nu=2/3$.

\subsection{Discrete distributions}

We are now going to discuss the case where the energies of the bonds take
discrete values. In this case, it may happen in Eq.~\eqref{DPrec} that
there are several paths coming from different potential parent
sites in the previous time section with the same minimal energy, and the
question is, of course, which path should be selected as the parent site.

\begin{figure}[!ht]
\centering
\includegraphics[width=.8\columnwidth]{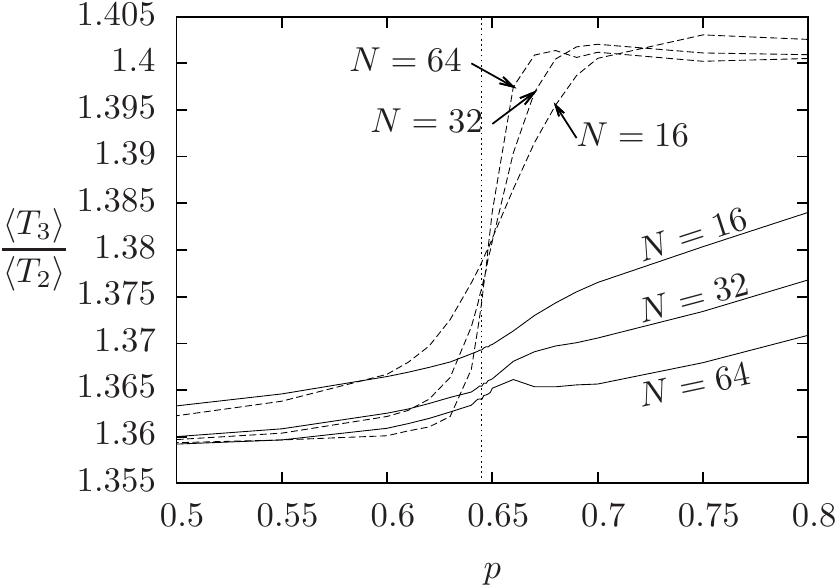}
\caption{Ratios $\langle T_3\rangle/\langle T_2\rangle$ as a function of
$p$ for the distribution of $\epsilon$ of Eq.~\eqref{disnoi}.
The dashed lines correspond to the simplest procedure of choosing with
equal probabilities one of the potential parent sites realizing the minimal energy,
and the plain lines represent the
results using the weights $\Omega$ which corresponds to the
$T\to0^+$ limit of finite temperature directed polymers.
The vertical dotted line indicates the directed percolation threshold on the
same lattice.}
\label{figdisc}
\end{figure}

The simplest idea is to choose randomly
at each time step with equal probabilities one of the paths with the
lowest energy. With this procedure, we have run numerical simulations in dimension $1+1$
for several sizes with a binary noise for the energies $\epsilon$ of the
bonds,
\begin{equation}
\epsilon=\begin{cases}	0 &\text{with probability $p$,}\\
			1 &\text{with probability $1-p$},
\end{cases}
\label{disnoi}
\end{equation}
for several values of $p$. Our results for the ratio $\langle
T_3\rangle/\langle T_2\rangle$ as a function of $p$ are shown in 
Fig.~\ref{figdisc} as dotted lines. 
As $p$ varies, we observe a crossover between two values: for small $p$,
$\langle
T_3\rangle/\langle T_2\rangle\approx 1.36$ as for directed polymers in
dimension $1+1$ when the distribution of energies is continuous (see
Fig.~\ref{ratios_DPT0}) and, for large $p$, $\langle
T_3\rangle/\langle T_2\rangle\approx 1.4$ which corresponds to the
coalescence of random walks in dimension~1, as in Eq.~\eqref{1d}.
The crossover between the two regions becomes sharper as $L$ increases,
which suggests a phase transition. The critical value of $p$ is very
consistent with the known threshold 0.6447 for directed percolation on
the same lattice\cite{Jensen.96}. Thus, the system
behaves like the neutral model when the $\epsilon=0$ bonds percolate.

Instead of choosing with equal probabilities which bond the
polymer follows when they are energetically equivalent, there is an
alternative procedure which corresponds to taking the limit $T\to0^+$ in
the problem of directed polymers at a finite temperature $T$.
At finite temperature, we keep track for each site
$A$ of the partition function $Z_A$ of a polymer arriving on $A$.
Assuming that the site $A$ has $M=2$ potential parent sites $B$ and $C$,
we have the recursion Eq.~(\ref{DPrec})
\def\of#1#2{_{#1\leftarrow#2}}
\begin{equation}
Z_A=Z\of{A}{B}+Z\of{A}{C},
\label{sumZ}
\end{equation}
where $Z\of{A}{B}=Z_B\exp(-\beta \epsilon_{(AB)})$ is the partition
function of a directed polymer arriving on $A$ via the site $B$ and where
$\beta=1/T$.
The probability that a polymer reaching $A$ comes from $B$ is given by
\begin{equation}
(\text{Prob. the polymer comes from $B$})={Z\of{A}{B}\over
Z_A}.
\label{probZ}
\end{equation}
At very low temperature, the partition function is dominated by the
lowest energy paths:
\begin{equation}
Z\approx\Omega e^{-\beta E},
\label{ZOE}
\end{equation}
where $E$ is the minimal energy and $\Omega$ the number of ways that this
energy $E$ can be obtained, so that Eq.~\eqref{sumZ} reads, at low
temperature,
\begin{equation}
\Omega_A e^{-\beta E_A} \approx \Omega_{B} e^{-\beta
E\of{A}{B}}+\Omega_{C} e^{-\beta E\of{A}{C}}.
\label{ZlT}
\end{equation}
where $E\of{A}{B}=E_B+\epsilon_{(AB)}$ is the minimal energy of the path
arriving at $A$ through $B$.
If $E\of{A}{B} < E\of{A}{C}$, then the first term in the right hand side of
Eq.~\eqref{ZlT} dominates and we obtain $E_A=E\of{A}{B}$ and
$\Omega_A=\Omega_B$. Furthermore, from Eq.~\eqref{probZ}, the
chosen path comes from $B$. On the other hand, if $E\of{A}{B} =
E\of{A}{C}$,
both terms in Eq.~\eqref{ZlT} have the same order of magnitude and we
obtain $E_A=E\of{A}{B}=E\of{A}{C}$ and $\Omega_A=\Omega_B+\Omega_C$.
Then, from Eq.~\eqref{probZ}, the probability that the directed
polymer comes from $B$ is $\Omega_B/\Omega_A$.

In this way, we not only choose the optimal energy but we also keep track of entropy effects.
We have run numerical simulations with the same parameters as above but
with this new procedure. The ratios $\langle
T_3\rangle/\langle T_2\rangle$ are shown on Fig.~\ref{figdisc} in plain
lines. For small values of $p$, both procedures yield the same results.
For larger $p$, however, the difference is striking, and the phase
transition seems to have disappeared: on both sides of the percolation
threshold the data seem to be in the same universality class (as they
converge to $\approx1.36$).

%\section{Models of population with selection}
%\label{sec:selection}

\section{Conclusion}

In this paper, we have presented analytical and numerical results showing the existence
of universality classes in the tree structures which appear in several models
of evolution and in directed polymers (see Tab.~\ref{Summary} for a summary).

Without selection, the genealogies of neutral models like the
Wright-Fisher model or coalescing random walks are described above the
critical dimension $d_c=2$ by the Kingman coalescent. For $d=1$ the
universality class is different: we have obtained the distribution
Eq.~(\ref{Tp1d}) of the ages $T_p$ of the most recent
common ancestor of $p$ individuals.

For directed polymers in a random medium, the same coalescence times
$T_p$ have been measured numerically. In the mean field case, their
values are compatible
with Bolthausen-Sznitman's coalescent, which is already known to appear
in spin glasses \cite{bolthausensznitman} and in branching random walks
with a selection mechanism keeping the size constant \cite{BDMM2.06,BDMM.07}. In low
dimension (at least $d=1$ and $d=2$), the coalescence times belong to
different
universality classes. It would be interesting to predict analytically
the numerical
values of $\langle T_3\rangle/\langle T_2\rangle$ and $\langle
T_4\rangle/\langle T_2\rangle$ measured in
Fig.~\ref{ratios_DPT0} for fast decaying distributions of $\epsilon$ as
well as the ones obtained in Fig.~\ref{fig1tail} for power-law
distributions of $\epsilon$ with exponent $\alpha\simeq 1^+$. In the
mean-field case, it would also be interesting to know if the replica method
can be used in order to determine the coalescence times.

The simulations presented in this paper deal only with directed polymers
at $T=0$. Directed polymers exhibit a phase transition for $d>2$ as the
temperature increases\cite{MonthusGarel.06}. We expect the tree
statistics to change at $T_c$ from the universality class of directed
polymers at zero temperature to the universality class of coalescing
random walks.

The construction of the minimal energy path for directed polymers can be
related to spatial models in presence of selection. In population
dynamics, selection can be taken into account through a parameter,
called the fitness or the adaptability, which characterizes the ability
of an individual to survive and reproduce\cite{Snyder.03, Kloster.04,
Kloster.05, Antal.07, DerridaSimon.07}. Individuals with a higher
fitness have a higher probability of having a descendance. This parameter
is transmitted from parents to offspring up to fluctuations due to
mutations. An analogy can be drawn between the minimal energy of a
directed polymer arriving on a site, and minus the fitness of an
individual living on a site. In presence of local selection, a spatial
model of population could therefore be formulated as follows: on each
site there would be one (or a finite number $m$ of individuals); at each
generation, each individual would branch into $k$ offspring with mutated
fitnesses. These offspring diffuse and, under the effect of selection,
only the best (or the $m$ best) individual(s) on each site would be kept.
Because of the similarity of such spatial models of population dynamics
in presence of selection with the directed polymers, we expect these
models to belong to the same universality classes.

We performed preliminary simulations on such a spatial model of evolution
with selection in dimension $1+1$  with $m=5$ individuals per site. Our
results for the ratios  $\langle T_3\rangle/\langle T_2\rangle$ and
$\langle T_4\rangle/\langle T_2\rangle$ coincide with those of directed
polymers.

\begin{table*}[!h]
\newcommand{\fraction}[2]{$\displaystyle{\strut#1\over\strut#2}$}
\begin{tabular}{|m{6cm}||c|c|c|c|c||c|} \hline
&\fraction{\langle T_3\rangle}{\langle T_2\rangle}
&\fraction{\langle T_4\rangle}{\langle T_2\rangle}
&\fraction{\langle T_N\rangle}{\langle T_2\rangle}
&\fraction{\langle T_2^2\rangle}{\langle T_2\rangle^2}
&\fraction{\langle T_3^2\rangle}{\langle T_2\rangle^2}
&$\langle T_2\rangle$ \\
\hline\hline
Kingman coalescent 
	& \multirow{4}{*}{\fraction{4}{3}}
	& \multirow{4}{*}{\fraction{3}{2}}
	& \multirow{4}{*}{2}
	& \multirow{4}{*}{2}
	& \multirow{4}{*}{\fraction{26}{9}}&$1$ \\[.3ex]
Coalescing random walks in $d>2$ 
	&&&&&&$\propto N$\\[.3ex]
Neutral models of evolution in $d>2$ 
	&&&&&&$\propto N$\\[.3ex]
Neutral model in $d=2$
	&&&&&&$\propto N\ln  N$\\
\hline
Neutral model in $d=1$
	&\fraction{7}{5}
	&\fraction{8}{5}
	&2
	&\fraction{12}{5}
	&\fraction{124}{35}
	&$N^2$\\
\hline
Bolthausen-Sznitman's coalescent 
	& \multirow{4}{*}{\fraction{5}{4}}
	& \multirow{4}{*}{\fraction{25}{18}}
	& \multirow{4}{*}{$\infty$}
	& \multirow{4}{*}{2}
	& \multirow{4}{*}{\fraction{11}{4}}
	&$1$ \\[.3ex]
Models of evolution with selection (or directed polymers at zero
temperature in mean field)
	&&&&&&$\propto (\ln N)^3$\\
\hline
Models of evolution with selection in dimension 2 (or directed polymers at
zero temperature in dimension $2+1$)
	& $\approx 1.29$
	& $\approx 1.42$
	& $\approx 1.93$
	& ? & ? & $\propto N^{1/(2\nu_{2+1})}\approx N^{0.80}$ \\
\hline
Models of evolution with selection in dimension 1 (or directed polymers at
zero temperature in dimension $1+1$)
	& $\approx 1.36$
	& $\approx 1.55$
	& $\approx 1.93$
	& ? & ? & $\propto N^{3/2}$ \\
\hline	
\end{tabular}
\caption{Universal ratios and order of magnitudes of coalescence times
for models of evolution with and without selection, and directed polymers
in a random medium. We could not reach large enough system sizes to give
a reliable numerical prediction for the ratios $\langle
T_p^2\rangle/\langle T_p\rangle^2.$}
\label{Summary}
\end{table*}

\bibliographystyle{apsrev}
\bibliography{journ,kpz,front,misc,treebib}

\end{document}